\documentclass{elsart}
\usepackage{amsmath}
\usepackage{epsfig}

\begin{document}

\begin{frontmatter}
\title{Two-Dimensional Lattice Boltzmann Model For Compressible Flows With
High Mach Number }
\author{Yanbiao Gan$^{1}$, Aiguo Xu$^{2*}$, Guangcai Zhang$^{2}$, Xijun Yu$^{2}$, Yingjun Li$^{1}$ }
\address{1, China University of Mining and Technology, Beijing
100083, P.R.China\\
2, National Key Laboratory of Computational Physics,\\
 Institute of
Applied Physics and Computational Mathematics, P.O.Box
8009-26, Beijing 100088, P.R.China\\
* Corresponding author. E-mail address: Xu\_Aiguo@iapcm.ac.cn
 }

\maketitle

\begin{abstract}

In this paper we present an improved lattice Boltzmann model for
compressible Navier-Stokes system with high Mach number. The model
is composed of three components: (i) the discrete-velocity-model by
Watari and Tsutahara [Phys Rev E \textbf{67},036306(2003)], (ii) a
modified Lax-Wendroff finite difference scheme where reasonable
dissipation and dispersion are naturally included,  (iii) artificial
viscosity. The improved model is convenient to compromise the high
accuracy and stability. The included dispersion term can effectively
reduce the numerical oscillation at discontinuity.  The added
artificial viscosity helps the scheme to satisfy the von Neumann
stability condition. Shock tubes and shock reflections are used to
validate the new scheme. In our numerical tests the Mach numbers are
successfully increased up to 20 or higher. The flexibility of the
new model makes it suitable for tracking shock waves with high
accuracy and for investigating nonlinear nonequilibrium complex
systems.
\end{abstract}

\end{frontmatter}

\section{Introduction}

Recently, the lattice Boltzmann(LB) method got substantial progress and has
been regarded as a promising alternative for simulating many complex
phenomena in various fields\cite{succi_book}. Unlike the macroscopic
computational fluid dynamics or the microscopic molecular dynamics, the LB
uses a mesoscopic discrete Boltzmann equation to describe the fluid system.
Because of its intrinsic kinetic nature, the LB contains more physical
connotation than Navier-Stokes or Euler equations based on the continuum
hypothesis\cite{dsmc_book}. From the Chapmann-Enskog analysis, the latter
can be derived from the former under the hydrodynamic limit.

Although having achieved great success in simulating incompressible fluids,
the application of LB to high-speed compressible flows still needs
substantial effort. High-speed compressible flows are ubiquitous in various
fields, such as explosion physics, aeronautics and so on\cite{XPZZ}.
Simulation of the compressible Navier-Stokes system, especially for the
those containing shock waves or contact discontinuities, is an interesting
and challenging work. Along the line, extensive efforts have been made in
the past years. Alexander, et al\cite{alex_pre_1992} presented a model where
the sound speed is selectable; Yan, et al\cite{ygw_pre_1999} proposed a
compressible LB model with three-speed-three-energy-level for the Euler
system; Yu and Zhao\cite{yhd_pre_2000} composed a model for compressible
flows by introducing an attractive force to soften sound speed; Sun\cite%
{sch_pre_1998,sch_pre_2000,sch_pre_2003} contributed a locally adaptive
semi-discrete LB model, where the set of particle speed is chosen according
to the local fluid velocity and internal energy so that the fluid velocity
is no longer limited by the particle speed set. In the development of LB for
Navier-Stokes systems, another way is referred to as the finite difference
lattice Boltzmann method (FDLBM)\cite{Tsutahara,watari_pre_2003,xag2005}.
The one by Watari-Tsutahara (WT) is typical \cite{watari_pre_2003}. The same
idea was then extended to binary compressible flows\cite{xag2005}. FDLBM\cite%
{watari_pre_2003,xag2005} breaks the binding of discretizations of space and
time and makes the particle speeds more flexible. But similar to previous LB
models, the numerical stability problem remains one of the few blocks for
its practical simulation to high Mach number flows. The stability problem of
LB has been addressed and attempted for some years \cite%
{lbe-1,Yong2003,Xiong2002,Tosi2006,Ansumali2002,Li2004,Sofonea2004,Brownlee2007,Seta,PXZJ,noi}
. Among them, the entropic LB method\cite{Tosi2006,Ansumali2002} tries to
make the scheme to follow the $H$-theorem; The FIX-UP method\cite%
{Tosi2006,Li2004} is based on the standard BGK scheme, uses a third order
equilibrium distribution function and a self-adapting updating parameter to
avoid negativeness of the mass distribution function. Flux limiter
techniques are used to enhance the stability of FDLB by Sofonea, et al\cite%
{Sofonea2004}. Adding minimal dissipation locally to improve stability is
also suggested by Brownlee, et al\cite{Brownlee2007}, but there such an
approach is not explicitly discussed. All the above mentioned attempts are
for low Mach number flows. In this paper we present a new LB scheme for
high-speed compressible flows which is composed of three components, (i) the
original DVM by WT, (ii) an Modified Lax-Wendroff (MLW) finite difference
scheme where reasonable dissipation and dispersion are naturally included,
(iii) additional artificial viscosity. With the new scheme, high speed
compressible flows with strong shocks can be successfully simulated.

This paper is organized as follows. In section 2 the original DVM by WT is
briefly reviewed. An alternative FD scheme combined with artificial
viscosity is introduced in section 3. The von Neumann stability analysis is
performed in section 4, from which solutions to improve the numerical
stability can be found. Several benchmark tests are used to validate the
proposed scheme in section 5. Section 6 presents the concluding remarks.

\section{Outline of the DVM by Watari-Tsutahara}

DVM of WT can be write as:
\begin{equation}
\mathbf{v}_{0}=0,\mathbf{v}_{ki}=v_{k}[\cos (\frac{i\pi }{4}),\sin (\frac{%
i\pi }{4})],i=1,2...8\text{,}  \label{dvm_eq_1}
\end{equation}%
where subscript $k$ indicates the $k$-th group of the particle velocities
whose speed is $v_{k}$ and $i$ indicates the direction of particle's speed.
A sketch of the DVM is referred to Fig.1.
\begin{figure}[tbp]
\center\includegraphics*[width=0.67\textwidth]{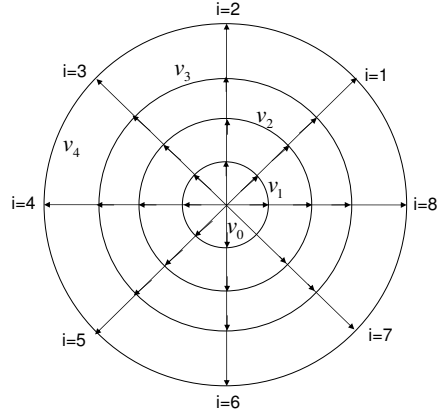}
\caption{ Sketch of the discrete-velocity-model used in the present paper.}
\end{figure}
It's easy to prove that this DVM at least up to seventh rank isotropy. The
evolution of the distribution function $f_{ki}$ with the
Bhatanger-Gross-Krook approximation\cite{bgk_1954} reads,
\begin{equation}
\frac{\partial f_{ki}}{\partial t}+\mathbf{v}_{ki}\cdot \frac{\partial f_{ki}%
}{\partial \mathbf{r}}=-\frac{1}{\tau }\left[ f_{ki}-f_{ki}^{eq}\right] ,
\label{bgk_eq}
\end{equation}%
where $f_{ki}^{eq}$ is the discrete version of the local equilibrium
distribution function; $\mathbf{r}$ is the spatial coordinate; $\tau $ is
the relaxation time; the local particle density $n$, hydrodynamic velocity $%
\mathbf{u}$ and temperature $T$ are defined by
\begin{equation}
n=\sum_{ki}f_{ki}^{eq},  \label{n_eq}
\end{equation}%
\begin{equation}
n\mathbf{u}=\sum_{ki}\mathbf{v}_{ki}f_{ki}^{eq},  \label{nu_eq}
\end{equation}%
\begin{equation}
P=e_{int}=nT=\sum_{ki}\frac{1}{2}(\mathbf{v}_{ki}-\mathbf{u})^{2}f_{ki}^{eq}
\label{p_eq}
\end{equation}%
where $P$ and $e_{int\text{ }}$are the local pressure and internal energy.
This model is designed to recover the following Navier-Stokes equations
\begin{equation}
\frac{\partial \rho }{\partial t}+\frac{\partial (\rho u_{\alpha })}{%
\partial r_{\alpha }}=0,  \label{ns_eq_1}
\end{equation}%
\begin{equation}
\frac{\partial (\rho u_{\alpha })}{\partial t}+\frac{\partial (\rho
u_{\alpha }u_{\beta }+P\delta _{\alpha \beta })}{\partial r_{\beta }}-\frac{%
\partial }{\partial r_{\beta }}[\mu (\frac{\partial u_{\beta }}{\partial
r_{\alpha }}+\frac{\partial u_{\alpha }}{\partial r_{\beta }}-\frac{\partial
u_{\gamma }}{\partial r_{\gamma }}\delta _{\alpha \beta })]=0,
\label{ns_eq_2}
\end{equation}%
\begin{gather}
\frac{\partial }{\partial t}[\rho (e_{int}+\frac{u^{2}}{2})]+\frac{\partial
}{\partial r_{\alpha }}[\rho u_{\alpha }(e_{int}+\frac{u^{2}}{2}+\frac{P}{%
\rho })]-\frac{\partial }{\partial r_{\alpha }}[\kappa ^{^{\prime }}\frac{%
\partial e_{int}}{\partial r_{\alpha }}  \notag \\
+\mu u_{\beta }(\frac{\partial u_{\beta }}{\partial r_{\alpha }}+\frac{%
\partial u_{\alpha }}{\partial r_{\beta }}-\frac{\partial u_{\gamma }}{%
\partial r_{\gamma }}\delta _{\alpha \beta })]=0  \label{ns3_eqq}
\end{gather}%
in the hydrodynamic limit, where $\mu $ ,$\kappa ^{^{\prime }}$ are
viscosity coefficient and heat conductivity coefficient, having the
following relations with pressure $P$ and relaxation time $\tau $ :%
\begin{equation}
\mu =P\tau ,\kappa ^{^{\prime }}=2\mu \text{.}  \label{miu_kapa_eq}
\end{equation}%
The equilibrium distribution function $f_{ki}^{eq}$ is calculated in the
following way,
\begin{eqnarray}
f_{ki}^{eq} &=&nF_{k}[(1-\frac{u^{2}}{2\theta }+\frac{u^{4}}{8\theta ^{2}})+%
\frac{v_{ki\varepsilon }u_{\varepsilon }}{\theta }(1-\frac{u^{2}}{2\theta })+%
\frac{v_{ki\varepsilon }v_{ki\pi }u_{\varepsilon }u_{\pi }}{2\theta ^{2}}(1-%
\frac{u^{2}}{2\theta })]+  \notag \\
&&\frac{v_{ki\varepsilon }v_{ki\pi }v_{ki\vartheta }u_{\varepsilon }u_{\pi
}u_{\vartheta }}{6\theta ^{3}}+\frac{v_{ki\varepsilon }v_{ki\pi
}v_{ki\vartheta }v_{ki\xi }u_{\varepsilon }u_{\pi }u_{\vartheta }u_{\xi }}{%
24\theta ^{4}}]  \label{feq_eq}
\end{eqnarray}%
with%
\begin{eqnarray}
F_{k} &=&\frac{1}{%
v_{k}^{2}(v_{k}^{2}-v_{k+1}^{2})(v_{k}^{2}-v_{k+2}^{2})(v_{k}^{2}-v_{k+3}^{2})%
}[48\theta ^{4}-6(v_{k+1}^{2}+v_{k+2}^{2}+v_{k+3}^{2})\theta ^{3}+  \notag \\
&&(v_{k+1}^{2}v_{k+2}^{2}+v_{k+2}^{2}v_{k+3}^{2}+v_{k+3}^{2}v_{k+1}^{2})%
\theta ^{2}-\frac{v_{k+1}^{2}v_{k+2}^{2}v_{k+3}^{2}}{4}\theta ]\text{,}
\label{fk_eq1}
\end{eqnarray}%
\begin{equation}
F_{0}=1-8(F_{1}+F_{2}+F_{3}+F_{4})\text{,}  \label{f0_eq}
\end{equation}%
and
\begin{equation*}
\theta =\frac{T}{m}\text{,}
\end{equation*}%
where
\begin{equation}
\left\{ k+l\right\} =\left\{
\begin{array}{ll}
k+l & \text{ if }k+l\leq 4 \\
k+l-4 & \text{ if }k+l>4%
\end{array}%
\right. \text{.}  \label{kl_eq}
\end{equation}%
We choose $v_{0}=0$ and four nonzero $v_{k}(k=1,2,3,4)$.

\section{Modified Lax-Wendroff scheme and artificial viscosity}

To simplify the discussion, we work on the general Cartesian coordinate. The
combination of the above DVM and the general FD scheme with first-order
forward in time and second-order upwinding in space composes the original
FDLB model by WT. It has been validated via the Couette flow, small Mach
number Riemann problems. When the Mach number $M$ exceeds $1,$ the original
LB model is not stable. The DVM is derived independent of Mach number.
Therefore, we resort to the discretization of the left-hand side of Eq. %
\eqref{bgk_eq} to make an accurate and stable LB scheme. Here we investigate
a mixed scheme which is composed of a modified Lax-Wendroff\cite{MLW} and an
artificial viscocity.

\bigskip As we know, the original Lax-Wendroff (LW) scheme is very
dissipative and has a strong \textquotedblleft smoothing effect". Obviously,
it is not favorable when needing capture shocks in the system. To compromise
the accuracy and stability, we add a dispersion term and the artificial
viscosity to the right-hand side of Eq. \eqref{bgk_eq} before discretization
so that we have
\begin{eqnarray}
\frac{\partial f_{ki}}{\partial t}+v_{ki\alpha }\frac{\partial f_{ki}}{%
\partial r_{\alpha }} &=&-\frac{1}{\tau }\left[ f_{ki}-f_{ki}^{eq}\right] +%
\frac{v_{ki\alpha }(1-c_{ki\alpha }^{2})\Delta r_{\alpha }^{2}}{6}\frac{%
\partial ^{3}f_{ki}}{\partial r_{\alpha }^{3}}  \notag \\
&&+\theta _{\alpha I}\left\vert \kappa _{\alpha }\right\vert (1-\left\vert
\kappa _{\alpha }\right\vert )\frac{\Delta r_{\alpha }^{2}}{2\Delta t}\frac{%
\partial ^{2}f_{ki}}{\partial r_{\alpha }^{2}}\text{,}  \label{MBe}
\end{eqnarray}%
where
\begin{equation}
c_{ki\alpha }=v_{ki\alpha }\Delta t/\Delta r_{\alpha }\text{, }\kappa
_{\alpha }=u_{\alpha }\Delta t/\Delta r_{\alpha }\text{; }  \label{add3}
\end{equation}%
\begin{equation}
\theta _{\alpha I}=\lambda \left\vert \frac{P_{\alpha I+1}-2P_{\alpha
I}+P_{\alpha I-1}}{P_{\alpha I+1}+2P_{\alpha I}+P_{\alpha I-1}}\right\vert
\label{add4}
\end{equation}%
plays a role of the switching function, $\lambda $ is a coefficient
controlling the amplitude of the artificial viscosity. Using the
Lax-Wendroff to the left-hand side and central difference to the right-hand
side of Eq. \eqref{MBe} results in the following LB equation,
\begin{eqnarray}
f_{kiI}^{new} &=&f_{kiI}-\frac{c_{ki\alpha }}{2}(f_{kiI+1}-f_{kiI-1})-\frac{%
\Delta t}{\tau }\left[ f_{kiI}-f_{kiI}^{eq}\right]  \notag \\
&&+\frac{c_{ki\alpha }^{2}}{2}(f_{kiI+1}-2f_{kiI}+f_{kiI-1})  \notag \\
&&+\frac{c_{ki\alpha }(1-c_{ki\alpha }^{2})}{12}%
(f_{kiI+2}-2f_{kiI+1}+2f_{kiI-1}-f_{kiI-2})  \notag \\
&&+\frac{\theta _{\alpha I}\left\vert \kappa _{\alpha }\right\vert
(1-\left\vert \kappa _{\alpha }\right\vert )}{2}%
(f_{kiI+1}-2f_{kiI}+f_{kiI-1})\text{,}  \label{dis_eq1}
\end{eqnarray}%
where the third suffixes $I-1,I,I+1$ indicate the mesh nodes in $x$ or $y$
direction. The positions of terms 3 and 4 in the right-hand side of Eq. %
\eqref{dis_eq1} have been exchanged. It is clear that the first line
corresponds to the general LB equation with the central difference in space;
compared with the central difference, the Lax-Wendroff contributes an extra
line II; lines III and IV show the added dispersion term and artificial
viscosity.

\section{von Neumann Stability Analysis}

We analysis the numerical stability of the FDLBM by means of von Neumann
stability analysis\cite{Seta,PXZJ}. In the analysis solution of
finite-difference equation is written as the familiar Fourier series, and
the numerical stability is evaluated by the magnitude of eigenvalues of an
amplification matrix. The small perturbation $\Delta f_{ki}$ is defined as: $%
f_{ki}(\mathbf{r},t)=\Delta f_{ki}(\mathbf{r},t)+\bar{f_{ki}^{0}}$, where $%
\bar{f_{ki}^{0}}$ is the global equilibrium distribution function which is a
constant, depends only on the mean density, velocity and temperature. From
equation\eqref{MBe} we can obtain
\begin{gather}
\frac{\Delta f_{ki}(r_{\alpha },t+\Delta t)-\Delta f_{ki}(r_{\alpha },t)}{%
\Delta t}+v_{ki\alpha }\frac{\partial \Delta f_{ki}}{\partial r_{\alpha }}=-%
\frac{1}{\tau }\left[ \Delta f_{ki}-\Delta f_{ki}^{eq}\right]  \notag \\
+\frac{v_{ki\alpha }(1-c_{ki\alpha }^{2})\Delta r_{\alpha }^{2}}{6}\frac{%
\partial ^{3}\Delta f_{ki}}{\partial r_{\alpha }^{3}}+\theta _{\alpha
I}\left\vert \kappa _{\alpha }\right\vert (1-\left\vert \kappa _{\alpha
}\right\vert )\frac{\Delta r_{\alpha }^{2}}{2\Delta t}\frac{\partial
^{2}\Delta f_{ki}}{\partial r_{\alpha }^{2}}\text{.}  \label{delt_fki_eq}
\end{gather}%
The perturbation part $\Delta f_{ki}(r_{\alpha },t)$ may be written as
series of complex exponents, $\Delta f_{ki}(r_{\alpha },t)=F_{ki}^{t}\mathrm{%
exp}(\mathbf{i}k_{\alpha }r_{\alpha })$, where $F_{ki}^{t}$ is an amplitude
at grid point $r_{\alpha }$ and time $t$, $\mathbf{i}$ is an imaginary unit,
and $k_{\alpha }$ is the wave number of sine wave in the domain with the
highest resolution $1/\Delta r_{\alpha }$. Substituting this expansion into
the equation \eqref{delt_fki_eq}, we obtain $F_{ki}^{t+\Delta
t}=G_{ij}F_{kj}^{t}$ , where $G_{ij}$ is a matrix being used to assess
amplification rate of $F_{ki}^{t}$ per time step $\Delta t$. If the maximum
of the eigenvalues of the amplification matrix satisfies the condition, $%
\mathrm{max}|\omega |\leq 1$, for all wave numbers, the FD scheme is surely
stable, where $\omega $ is the eigenvalue of the amplification matrix. This
is the von Neumann condition for stability. The amplification matrix $G_{ij}$
can be written as following,%
\begin{align}
G_{ij}& =\left( 1-\frac{\Delta t}{\tau }\right) \delta _{ij}+\frac{\Delta t}{%
\tau }\frac{\partial f_{ki}^{eq}}{\partial f_{kj}}-\frac{c_{ki\alpha }}{2}%
(e^{\mathbf{i}k_{\alpha }\Delta r_{\alpha }}-e^{-\mathbf{i}k_{\alpha }\Delta
r_{\alpha }})\delta _{ij}+\frac{c_{ki\alpha }^{2}}{2}(e^{\mathbf{i}k_{\alpha
}\Delta r_{\alpha }}-2  \notag \\
& +e^{-\mathbf{i}k_{\alpha }\Delta r_{\alpha }})\delta _{ij}+\frac{%
c_{ki\alpha }(1-c_{ki\alpha }^{2})}{12}(e^{\mathbf{i}2k_{\alpha }\Delta
r_{\alpha }}-2e^{\mathbf{i}k_{\alpha }\Delta r_{\alpha }}+2e^{-\mathbf{i}%
k_{\alpha }\Delta r_{\alpha }}-e^{-\mathbf{i}2k_{\alpha }\Delta r_{\alpha
}})\delta _{ij}  \notag \\
& +\frac{\theta _{\alpha I}\left\vert \kappa _{\alpha }\right\vert
(1-\left\vert \kappa _{\alpha }\right\vert )}{2}(e^{\mathbf{i}k_{\alpha
}\Delta r_{\alpha }}-2+e^{-\mathbf{i}k_{\alpha }\Delta r_{\alpha }})\delta
_{ij}\text{,}  \label{gi_j_eq}
\end{align}%
\begin{equation}
\frac{\partial f_{ki}^{eq}}{\partial f_{kj}}=\frac{\partial f_{ki}^{eq}}{%
\partial \rho }\frac{\partial \rho }{\partial f_{kj}}+\frac{\partial
f_{ki}^{eq}}{\partial T}\frac{\partial T}{\partial f_{kj}}+\frac{\partial
f_{ki}^{eq}}{\partial u_{\alpha }}\frac{\partial u_{\alpha }}{\partial f_{kj}%
}\text{.}  \label{pian_fki_eq}
\end{equation}%
Several researchers have analyzed the stability of the incompressible LB
models\cite{Seta,jdsc_jcp,xd_jsp_2004}, it is found that there is not a
single wave-number being always the most unstable. For the 2D DVM by WT, $%
G_{ij}$ is a matrix with $33\times 33$ elements. Moreover, every element is
related to the macroscopical variables (density, temperature, velocities),
discrete velocities and other constants, so it is difficult to analyze with
explicit expressions. We resort to using the software, Mathematica-5 to
conduct a series of quantitative analysis. Now we show some numerical
results of von Neumann analysis by Mathematica-5. The results will be shown
by figures with curves for the maximum eigenvalue $|\omega |_{max}$ of $%
G_{ij}$ versus $k\Delta x$.

Figure 2 shows a comparison of the fours different cases, (i) LB with only
the central-difference to the convection term, i.e. LB scheme based only on
the first line of Eq. \eqref{dis_eq1} (see the black line with squares);
(ii) LB with Lax-Wendroff, i.e., LB scheme based on the first two lines of
Eq. \eqref{dis_eq1} (see the red line with circles); (iii) LB scheme based
on lines 1 and 3 of Eq. \eqref{dis_eq1} (see the green line with triangles);
(iv) LB scheme based on the whole of Eq. \eqref{dis_eq1} (see the blue line
with triangles down). Here the macroscopic variables are chosen as $(\rho
,u_{1},u_{2},T)$ = $(2.0,30.0,0.0,1.0)$, and the remaining parameters are $%
(v_{0},v_{1},v_{2},v_{3},v_{4})$ = $(0.0,1.0,1.92,2.99,4.49)$, $\lambda =0.5$%
, $\Delta x=\Delta y=3\times 10^{-3}$, $\Delta t=\tau =10^{-5}$. It is clear
that the artificial viscosity term can significantly decrease the maximum
eigenvalue $|\omega |_{\text{max}}$ from being larger than to be smaller
than $1$ for appropriately given time step. Moreover, it is worthy to note
that dissipation term in line 2 of Eq. \eqref{dis_eq1} favors and dispersion
term in line 3 disfavors the stability to some extent. Numerical experiments
show that the dispersion term may effectively reduce numerical oscillations
near discontinuity and improves the accuracy (see Fig. 3 for an example).
\begin{figure}[tbp]
\center\includegraphics*[width=0.67\textwidth]{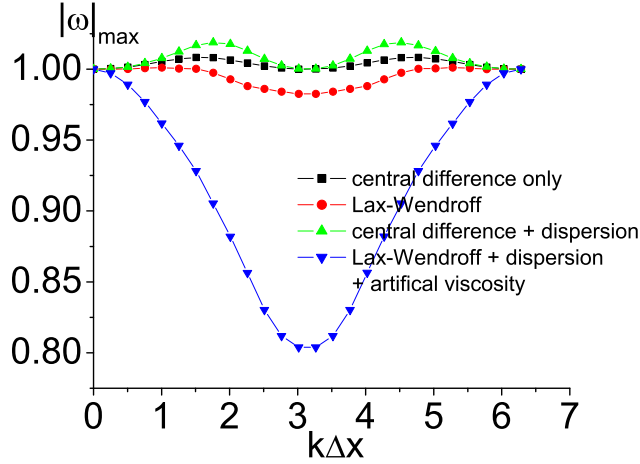}
\caption{ (Color online) Stability analysis for four conditions. The
macroscopic variables are set as $(\protect\rho ,u_{1},u_{2},T)$ = $%
(2.0,30.0,0.0,1.0)$, the other constants are set as $%
(v_{0},v_{1},v_{2},v_{3},v_{4})$ = $(0.0,1.0,1.92,2.99,4.49)$, $\protect%
\lambda =0.5$,$\Delta x=\Delta y=3\times 10^{-3}$,$\Delta t=$ $\protect\tau $
$=10^{-5}$.}
\end{figure}
\begin{figure}[tbp]
\center\includegraphics*[width=0.67\textwidth]{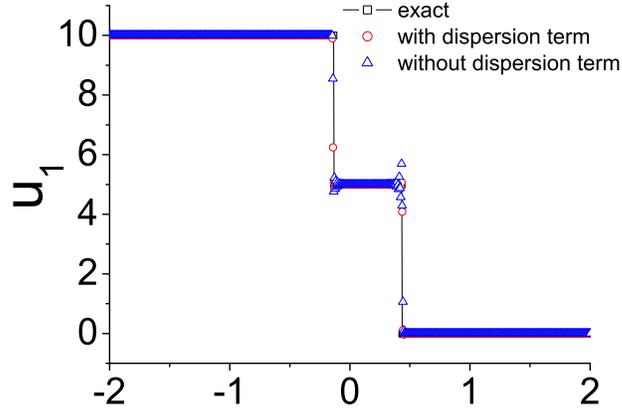}
\caption{(Color online) Effects of dispersion to simulation. The macroscopic
variables at two sides of a shock tube are set as $(\protect\rho ,
u_{1},u_{2},T)|_{L}=(10.0,10.0,0.0,5.0)$, $(\protect\rho %
,u_{1},u_{2},T)|_{R}=(10.0,0.0,0.0,5.0)$, $\protect\lambda =2.0$. The other
constants and macroscopic variables are the same as in Fig.2.}
\end{figure}

Figure 4 shows the effects of various artificial viscosities to the
stability. Fig.(a) shows the cases with $\lambda =1.0$, $0.5$, $0.1$, and $%
0.05$. Fig.(b) shows the cases with $\lambda =1.0$, $1.8$,$2.0$, and $3.0$.
The other constants and macroscopic variables are unchanged. From this
figure we can see some relevance: Strength of artificial viscosity has a
large impact on the stability. LB works only within a certain range of
artificial viscosity. In practical simulations, we generally take the
smaller viscosity in favor of the accuracy.
\begin{figure}[tbp]
\center\includegraphics*[width=0.9\textwidth]{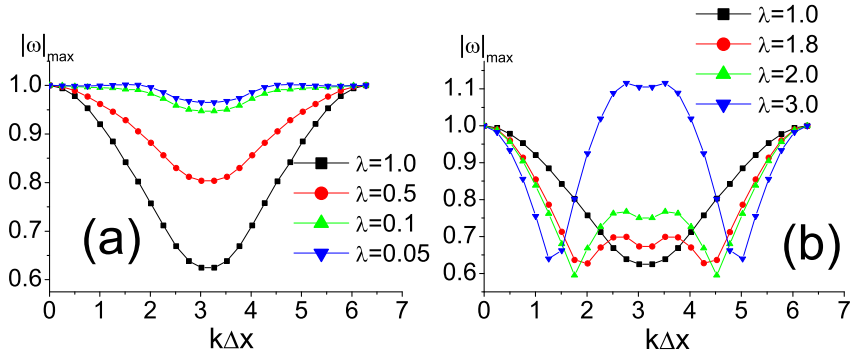}
\caption{(Color online) Effects of various artificial viscosities to the
numerical stability. Fig. (a) shows the cases with $\protect\lambda =1.0$, $%
0.5$, $0.1$, $0.05$. Fig. (b) shows the cases with $\protect\lambda =1.0$,$%
1.8$, $2.0$, $3.0$. The other constants and macroscopic variables are
unchanged.}
\end{figure}

Since the density $\rho $ can be normalized to $1$, we then need only
investigate the effects of the other two physical quantities, temperature $T$
and flow velocity $\mathbf{u}$. Figure 5 shows four cases with $T=1$, $T=5$,
$T=15$ and $T=25$. Here $u_{1}=5$, $u_{2}=0$ and $\lambda =0$. When other
parameters are fixed, the numerical stability becomes better with the
increasing of temperature. This can also be understood that higher
temperature corresponds to higher sound speed and lower Mach number.
\begin{figure}[tbp]
\center\includegraphics*[width=0.67\textwidth]{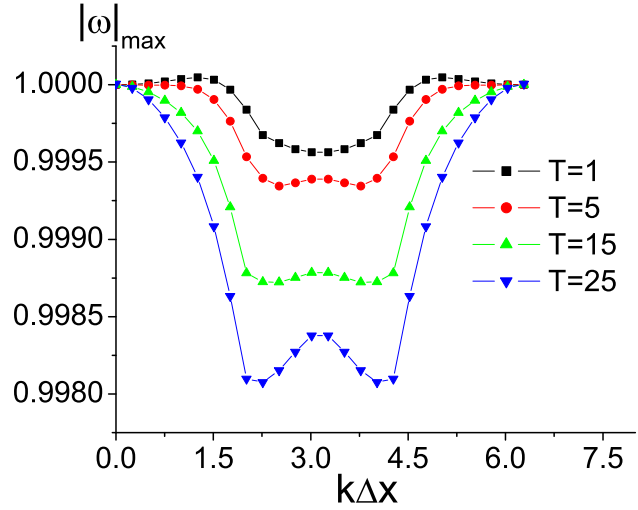}
\caption{(Color online) Influence of temperature $T$ to numerical stability.
$u_{1}=5$, $u_{2}=0$ and $\protect\lambda =0$. The other physical quantities
and model parameters are unchanged.}
\end{figure}
\begin{figure}[tbp]
\center\includegraphics*[width=0.67\textwidth]{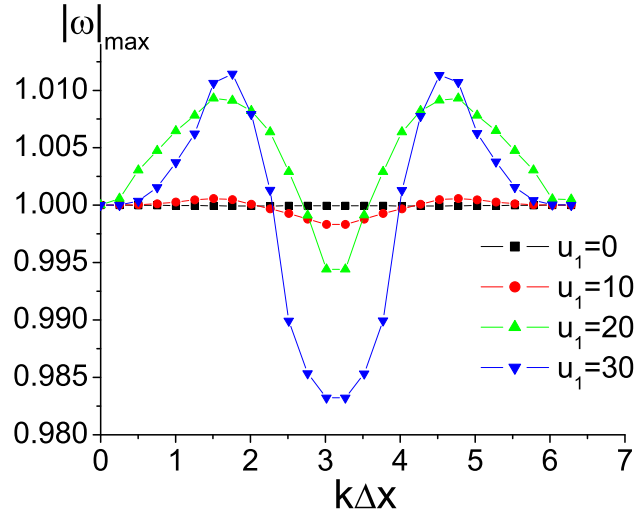}
\caption{(Color online) Influence of velocity $\mathbf{u}$ to numerical
stability. The value of $u_{1}$ is altered from zero to $30$ and $u_{2}=0$.
Here $\protect\lambda =0$, the other constants and macroscopic variables are
unchanged.}
\end{figure}

Figure 6 shows cases with difference flow velocities. The value of $u_{1}$
is altered from zero to $30$ and $u_{2}=0$. Here $\lambda =0$, the other
constants and macroscopic variables are unchanged. This figure clearly shows
that the higher the Mach number, the larger the maximum eigenvalue, which
answers why the numerical stability becomes worse with the increasing of
Mach number of the fluid.

\section{Numerical tests and analysis}

In this section two kinds of typical benchmarks are used to validate the
newly proposed scheme. The first kind is the Riemann problem\cite{Lv}. The
second one is the problem of shock reflection\cite{PhD}.

\subsection{Riemann problems\protect\cite{Lv}}

Here the two-dimensional model is used to solve the one-dimensional Riemann
problem. The initial macroscopic variables at the two sides are denoted by $%
(\rho $, $u_{1}$, $u_{2}$, $T)|_{L}$ and $(\rho $, $u_{1}$, $u_{2}$, $T)|_{R}
$, respectively.

\subsubsection{Sod's shock tube}

For the problem considered, the initial condition is described by
\begin{equation}
\left\{
\begin{array}{l}
(\rho ,u_{1},u_{2},T)|_{L}=(1.0,0.0,0.0,1.0) \\
(\rho ,u_{1},u_{2},T)|_{R}=(0.125,0.0,0.0,0.8)%
\end{array}%
\right.   \label{sod_eq}
\end{equation}%
Figure 7 shows the computed density, pressure, velocity, temperature
profiles at $t=0.2$, where the circles are simulation results and solid
lines with squares are analytical solutions. The size of grid is $\Delta
x=\Delta y=10^{-3}$, time step $\Delta t=10^{-5}$, and $\tau =$ $10^{-4}$, $%
\lambda =2.$ The two sets of results have a satisfying agreement.
\begin{figure}[tbp]
\center\includegraphics*[width=0.8\textwidth]{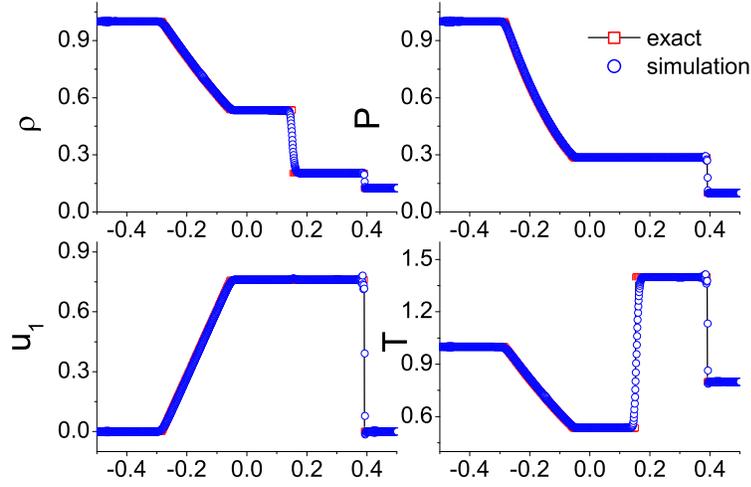}
\caption{(Color online) Comparison of numerical and theoretical results for
the Sod shock tube, where t=0.2. Solid lines with squares are for exact
solutions and circles are for simulation results.}
\end{figure}

\subsubsection{Lax's shock tube}

The initial condition of this problem reads
\begin{equation}
\left\{
\begin{array}{l}
(\rho ,u_{1},u_{2},T)|_{L}=(0.445,0.698,0.0,7.928) \\
(\rho ,u_{1},u_{2},T)|_{R}=(0.5,0.0,0.0,1.142)%
\end{array}%
\right.  \label{lax_eq}
\end{equation}%
Figure 8 shows the results at $t=0.2$, where the circles are simulation
results and solid lines with squares correspond to exact solutions. The
parameters are set to be $\Delta x=\Delta y=10^{-3}$, $\Delta t=10^{-5}$, $%
\tau =$ $10^{-4}$, and $\lambda =1$. We also find a good agreement between
the two sets of results.
\begin{figure}[tbp]
\center\includegraphics*[width=0.8\textwidth]{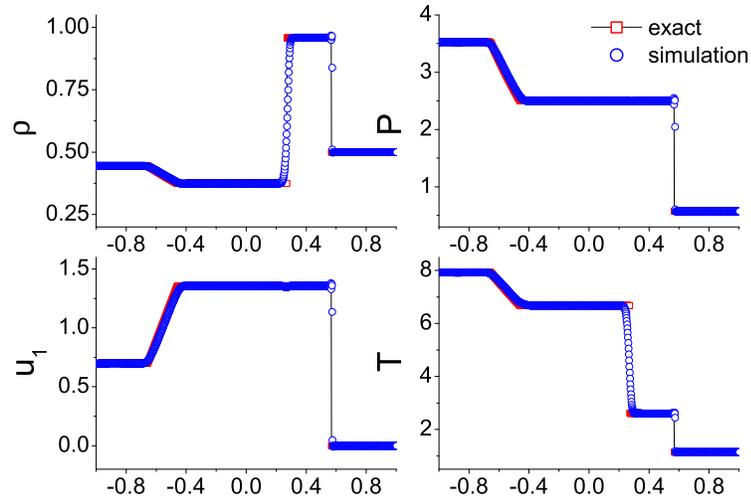}
\caption{(Color online) Comparison of numerical and theoretical results for
the Lax shock tube at $t=0.2$. Solid lines with squares are for exact
solution and circles are for simulation results.}
\end{figure}

\subsubsection{Sjogreen's problem}

The initial condition of this problem is
\begin{equation}
\left\{
\begin{array}{l}
(\rho ,u_{1},u_{2},T)|_{L}=(1.0,-2.0,0.0,0.4) \\
(\rho ,u_{1},u_{2},T)|_{R}=(1.0,2.0,0.0,0.4)%
\end{array}%
\right.  \label{sjo_eq}
\end{equation}%
The numerical and exact solutions at $t=0.018$ are shown in Fig.9, where $%
\Delta x=\Delta y=3\times 10^{-3}$, $\Delta t=\tau =10^{-5}$ and $\lambda
=1.8$. The exact solution of this problem consists of two strong rarefaction
waves and a weak constant contact discontinuity. Pressure near the contact
discontinuity is very small, which brings certain difficulties to
simulation. Temperature and density calculated by many schemes are negative.
However, the improved model ensures the positivity of them. Successful
simulation of this problem proves that the improved model is applicable to
the low-density, low-temperature flow simulations.
\begin{figure}[tbp]
\center\includegraphics*[width=0.8\textwidth]{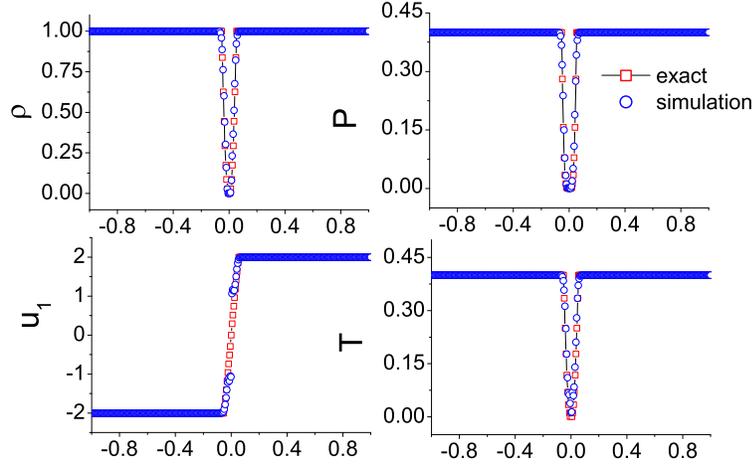}
\caption{(Color online) Comparison of numerical and theoretical results for
the Sjogreen problem at $t=0.018$. Solid lines with squares are for exact
solutions, and circles are for simulation results.}
\end{figure}

\subsubsection{Colella's explosion wave problem}

The initial condition of this test can be write as
\begin{equation}
\left\{
\begin{array}{l}
(\rho ,u_{1},u_{2},T)|_{L}=(1.0,0.0,0.0,1000.0) \\
(\rho ,u_{1},u_{2},T)|_{R}=(1.0,0.0,0.0,0.01)%
\end{array}%
\right.  \label{wood_eq}
\end{equation}%
This is generally regarded as a difficult test. The exact solution contains
a leftwards rarefaction wave, a contact discontinuity and a strong shock. It
is generally used to check the robustness and accuracy. Figure 10 gives
comparison of the numerical and theoretical results at $t=0.05$. Here $%
\lambda =20$, other parameters are same as in the Sjogreen test. Successful
simulation of this test proves the improved model is applicable to flows
with very high ratios of temperature and pressure.

\begin{figure}[tbp]
\center\includegraphics*[width=0.8\textwidth]{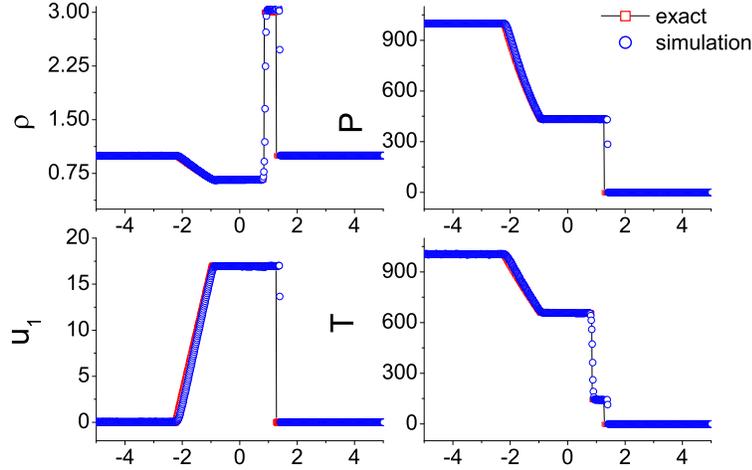}
\caption{(Color online) Comparison of numerical and theoretical results for
the Colella explosion wave problem at t=0.05. Solid lines with squares are
for exact solutions and circles are for simulation results.}
\end{figure}

\subsubsection{Collision of two strong shocks}

This test with the following initial data:
\begin{equation}
\left\{
\begin{array}{l}
(\rho ,u_{1},u_{2},T)|_{L}=(5.99924,19.5975,0.0,76.8254) \\
(\rho ,u_{1},u_{2},T)|_{R}=(5.99242,-6.19633,0.0,7.69222)%
\end{array}%
\right.  \label{colli_eq}
\end{equation}%
This is also a difficult test. Exact solution contains a leftwards shock, a
right contact discontinuity and shock which spreading to right side. And the
left-shock spreads to right very slowly, which brings additional
difficulties to the numerical method. Fig.11 gives a comparison of the
numerical and theoretical results at $t=0.12$. Parameters used in this test
are $\Delta x=\Delta y=2\times 10^{-3}$, $\Delta t=\tau =10^{-5}$ and $%
\lambda =1$. The good agreement between the two sets of results shows again
the robustness of the improved model.
\begin{figure}[tbp]
\center\includegraphics*[width=0.8\textwidth]{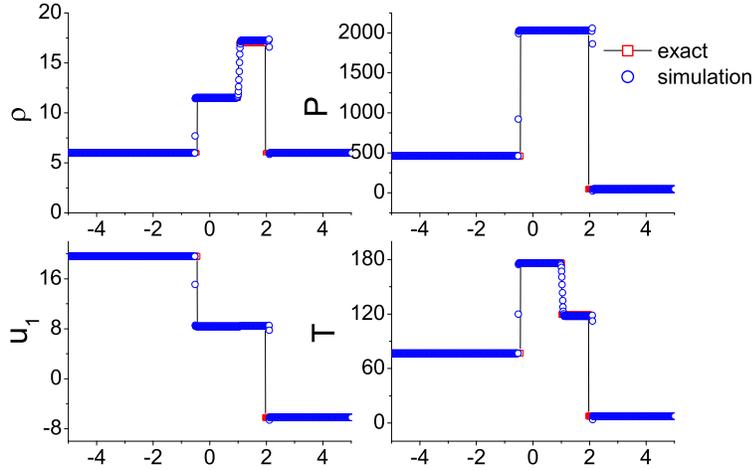}
\caption{(Color online) Comparison of numerical and theoretical results for
collision of two strong shocks at $t=0.12$. Solid lines with symbols are for
exact solutions and symbols are for simulation results.}
\end{figure}

\subsection{Shock reflections\protect\cite{PhD}}

We will present two gas dynamics simulations. Both are done on rectangular
grid. The first is to recover a steady regular shock reflection. The second
is the double Mach reflection of a shock off an oblique surface. This
example is used in Ref.\cite{Woodward1984} as a benchmark test for comparing
the performance of various difference methods on problem involving strong
shocks.

\subsubsection{Steady regular shock reflection}

In the first test problem, the incoming shock wave with Mach number 20 has
an angle of $30^{\circ }$ to the wall. The computational domain is a
rectangle with length $0.9$ and height $0.3$. This domain is divided into a $%
300\times 100$ rectangular grid with $\Delta x=\Delta y=0.003$. The boundary
conditions are composed of a reflecting surface along the bottom boundary,
supersonic outflow along the right boundary , and Dirichlet conditions on
the left and top boundary conditions, given by
\begin{equation}
\left\{
\begin{array}{l}
(\rho \text{, }u_{1}\text{, }u_{2}\text{, }T)|_{0\text{, }y\text{, }t}=(1.0%
\text{, }20.0\text{, }0.0\text{, }0.5)\text{,} \\
(\rho \text{, }u_{1}\text{, }u_{2}\text{, }T)|_{x\text{, }0.3\text{, }t}=(%
\frac{50}{17}\text{, }16.7\text{, }-5.71578\text{, }22.61)\text{.}%
\end{array}%
\right.  \label{zg_eq}
\end{equation}%
Figure 12 shows contours of density at $t=0.2$. The clear shock reflection
on the wall agrees well with the exact solution.
\begin{figure}[tbp]
\center\includegraphics*[width=0.8\textwidth]{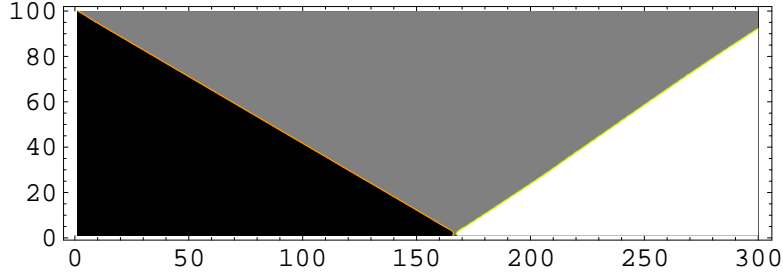}
\caption{(Color online)Density contour of steady regular shock reflection on
a wall at t=0.2. From black to white, the density increases.}
\end{figure}

\subsubsection{Double Mach reflection}

In this test, we considered an unsteady shock reflection. The initial
pressure ratio here is high. A planar shock is incident towards an oblique
surface with a $30^{\circ }$ angle to the direction of propagation of the
shock. A uniform mesh size of $500\times 200$ is used for the numerical
simulation. The conditions for both sides are:
\begin{figure}[tbp]
\center\includegraphics*[width=0.8\textwidth]{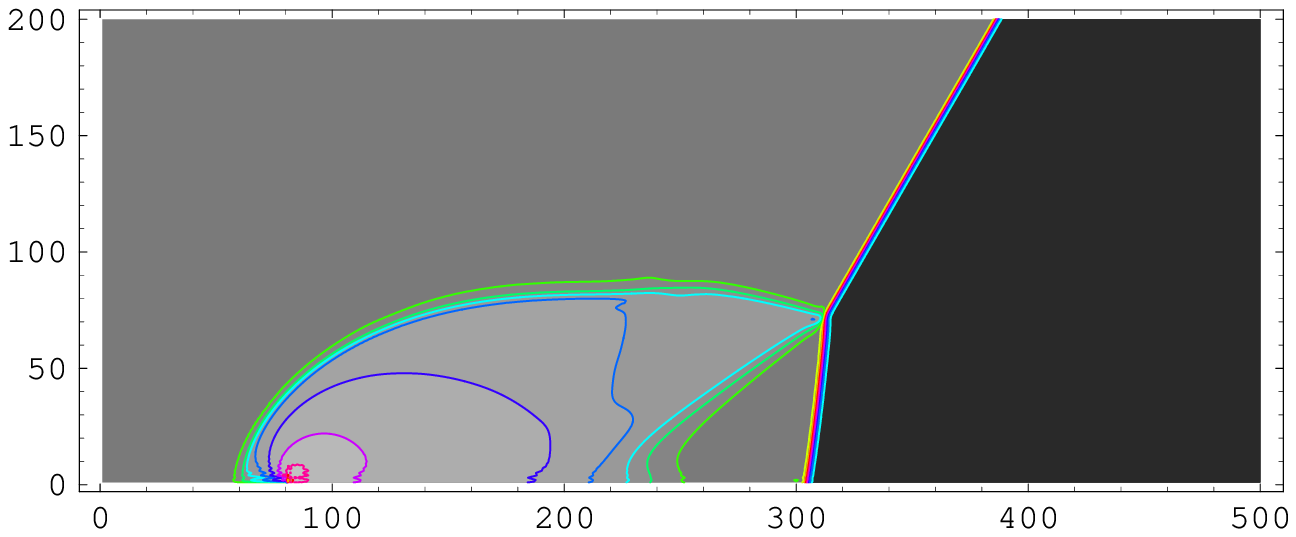} \center%
\includegraphics*[width=0.8\textwidth]{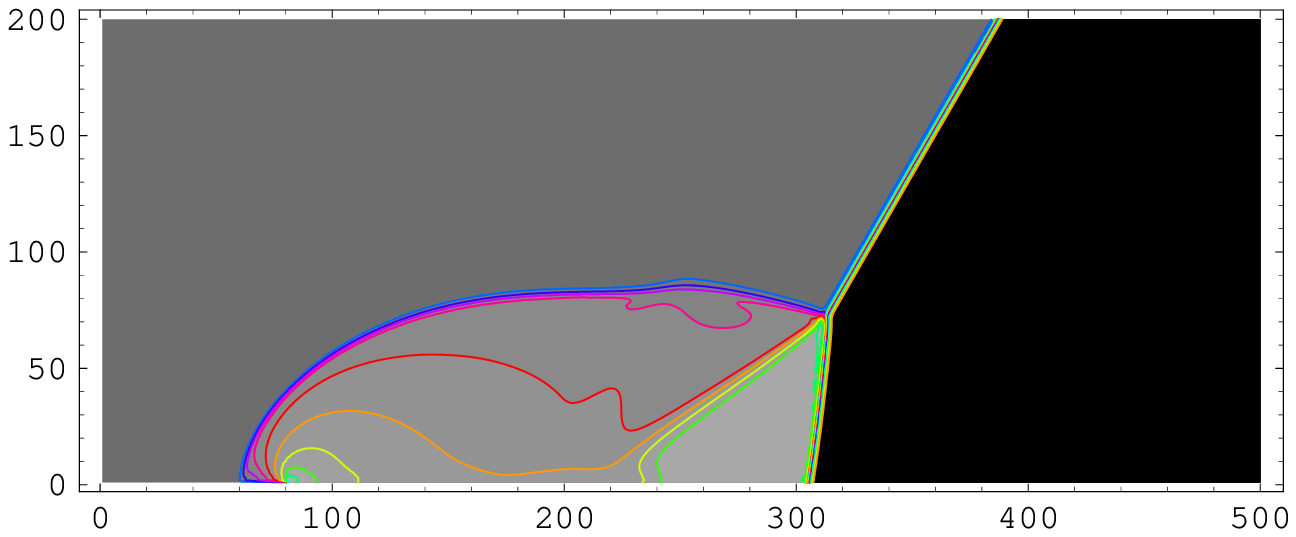} \center\includegraphics*%
[width=0.8\textwidth]{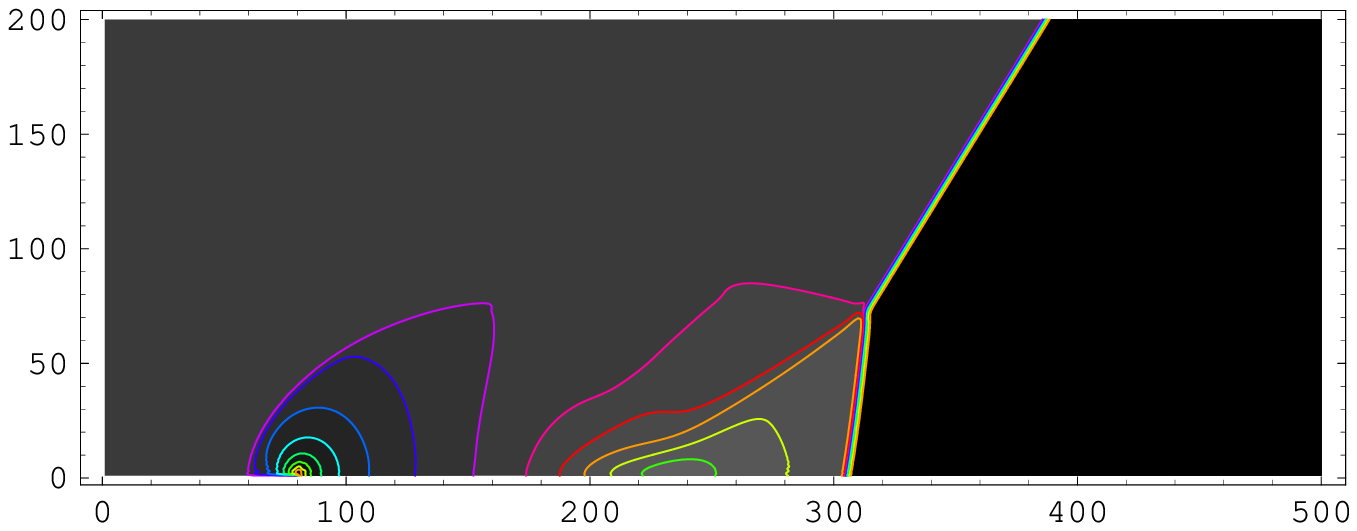}
\caption{(Color online) Contours of density (top), temperature (center), and
$u_{1}$ (bottom) of the double Mach reflection problem at the 750th
iteration step. The units of the $x$- and $y$- axes are both $0.001$. }
\end{figure}
\begin{equation}
\left( \rho ,u_{1},u_{2},T\right) \mid _{x\text{,}y\text{,}0}=\left\{
\begin{array}{ll}
(\frac{400}{67},13.3\cos 30^{\circ },-13.3\sin 30^{\circ },89.2775), & \text{
if }y\geq h(x,0) \\
(2.0,0.0,0.0,0.5) & \text{ if }y<h(x,0)%
\end{array}%
\right. \text{,}
\end{equation}%
where $h(x,t)=\sqrt{3}(x-80\Delta x)-40t$. The reflecting wall lines along
the bottom of the problem domain, beginning at $x=0.08.$ The shock makes a $%
60^{\circ }$ angle with the $x$ axis and extends to the top of the problem
domain at $y=0.2.$ At the top boundary, the physical quantities are assigned
the same values as on the left side for $x\leq g(t)$ and are assigned the
same values on the right side, where $g(t)=80\Delta x+\sqrt{3}/3(0.2+40t)$.
The computed density, temperature and flow velocity along the $x$-direction
are shown in Fig.13, where complex characteristics, such as oblique shocks
and triple points, are well captured.

\section{Conclusions and discussions}

A lattice Boltzmann model to the high-speed compressible Navier-Stokes
system is presented. The new LB model is composed of the following
components: the original DVM by Watari and Tsutahara, a modified
Lax-Wendroff scheme and an additional artificial viscosity. Compared with
the central difference scheme, the Lax-Wendroff contributes a dissipation
term which is in favor of the numerical stability, even though it is
generally still not enough for high-speed flows. The introducing of the
third-order dispersion term helps to eliminate some unphysical oscillations
at discontinuity. The additional artificial viscosity compensates the
insufficiency of the above-mentioned dissipation so that the LB simulation
can continue smoothly. The adding of the dispersion and artificial viscosity
terms should survive the dilemma of stability versus accuracy. In other
words, they should be minimal but make the evolution satisfy the von Neumann
stability condition. Due to the complexity, the analysis resorts to the
software, Mathematica-5, and only some typical results are shown by figures.

Typical benchmark tests are used to validate the proposed scheme. Riemann
problems, including the Sod, Lax, Sjogreen, Colella explosion wave,
collision of two strong shocks, show good accuracy and numerical stability
of the new scheme, even though they are generally difficult to resolve by
traditional computational fluid dynamics\cite{Lv,PhD,Woodward1984,yu xijun}.
Regular and double Mach shock reflections are successfully recovered. These
simulations show that the improved LB model may be used to investigate some
long-standing problems, such as the transitions between regular and Mach
reflections. By incorporating an appropriate equation of state, or
equivalently, a free energy functional, or an external force, the present
model may be used to simulate liquid-vapor transition and relevant flow
behavior. Future work includes a more complete description of the problem on
numerical accuracy versus stability and thermal lattice Boltzmann model for
multi-phase flows.

\section*{acknowledgments}

This work is supported by the National Basic Research Program (973 Program)
[under Grant No. 2007CB815105], Science Foundation of Laboratory of
Computational Physics, National Natural Science Foundation [under Grant Nos.
10775018,10474137 and 10702010] of China.

\end{document}